\documentclass[twoside]{elsart}

\usepackage{amsmath,epsfig,amsfonts,amssymb,latexsym}

\begin{document}
\date{\today}
\begin{frontmatter}
\title			{Basel II for Physicists: A Discussion Paper}
 
\author[Alessandria]	{Enrico Scalas}

\address[Alessandria]	{Dipartimento di Scienze e Tecnologie Avanzate, 
			Universit\`a del Piemonte Orientale, 
			Corso Borsalino 54, 
			I--15100 Alessandria, Italy}

\begin{abstract}
On June 26th, 2004, Central bank governors and the heads of bank supervisory authorities in the Group of Ten (G10) countries issued a press release and endorsed the publication of {\em International Convergence of Capital Measurement and Capital Standards: a Revised Framework}, the new capital adequacy framework commonly known as Basel II. According to Jean Claude Trichet, Chairman of the G10 group of central bank governors and heads of bank supervisory authorities and President of the European Central Bank: ``Basel II embraces a comprehensive approach to risk management and bank supervision. It will enhance banks' safety and soundness, strengthen the stability of the financial system as a whole, and improve the financial sector's ability to serve as a source for sustainable growth for the broader economy.'' The negotial process is likely to lead to the
adoption of the new rules within 2007. In 1996, after the {\em Amendment to the capital accord to incorporate market risks}, a new wave of physicists entered risk
management offices of large banks, that had to develop internal models of
{\em market risk}. Which will be the challenges and opportunities for physicists
in the financial sector in the years to come? This paper is a first modest contribution for starting a debate within the Econophysics community.
\end{abstract}

\begin{keyword}
finance, statistical finance
\\ {{\it JEL: \ }} G21; G28
\\ {\it Corresponding author}: Enrico Scalas ({\tt scalas@unipmn.it}), 
\\ url: {\tt www.econophysics.org}
\end{keyword}

\end{frontmatter}

\def\eg{{\it e.g.}\ } \def\ie{{\it i.e.}\ }
\def\sg{\hbox{sign}\,}
\def\sgn{\hbox{sign}\,}
\def\sign{\hbox{sign}\,}
\def\e{\hbox{e}}
\def\exp{\hbox{exp}}
\def\ds{\displaystyle}
\def\dis{\displaystyle}
\def\q{\quad}    \def\qq{\qquad}
\def\lan{\langle}\def\ran{\rangle}
\def\l{\left} \def\r{\right}
\def\lra{\Longleftrightarrow}
\def\arg{\hbox{\rm arg}}
\def\d{\partial}
 \def\dr{\partial r}  \def\dt{\partial t}
\def\dx{\partial x}   \def\dy{\partial y}  \def\dz{\partial z}
\def\rec#1{{1\over{#1}}}
\def\log{\hbox{\rm log}\,}
\def\erf{\hbox{\rm erf}\,}     \def\erfc{\hbox{\rm erfc}\,}
\def\F{\hbox{F}\,}
\def\NN{\hbox{\bf N}}
\def\RR{\hbox{\bf R}}
\def\CC{\hbox{\bf C}}
\def\ZZ{\hbox{\bf Z}}
\def\II{\hbox{\bf I}}


\section{Introduction}

On June 26th, 2004, Central bank governors and the heads of bank supervisory authorities in the Group of Ten (G10) countries issued a press release and endorsed the publication of {\em International Convergence of Capital Measurement and Capital Standards: a Revised Framework} \cite{basel2}, the new capital adequacy framework commonly known as Basel II. The negotial process is likely to lead to the adoption of the new rules within 2007. In 1996, after the {\em Amendment to the capital accord to incorporate market risks}, a new wave of physicists was included in risk management offices of large banks, that had to develop internal models of {\em market risk}. What challenges and opportunities 
will there be for physicists working in the financial sector in the years to come? This paper is a first modest contribution to a start a debate within the Econophysics community. It is divided into three parts. In section 2, the reader will find a short description of Basel II pillars. Section 3 will be devoted to analyse some of the possible subjects on which physicists could provide new and interesting insights. Finally, in section 4, preliminary conclusions for the European research and educational systems will be drawn.

\section{The Basel II framework}

In order to quickly present the framework, it is useful to make reference to the press release of June 26th 2004 \cite{release}. The text taken from the press release, will be
included within double brakets $\langle \langle$ ... $\rangle \rangle$. The press release does not address critical and controversial issues, some of which will be mentioned below.

\subsection{Some history}

$\langle \langle$ The 1988 Basel Capital Accord set out the first internationally accepted definition of, and a minimum measure for, bank capital. The Basel Committee designed the 1988 Accord as a simple standard so that it could be applied to many banks in many jurisdictions. It requires banks to divide their exposures up into broad {\em classes} reflecting similar types of borrowers. Exposures to the same kind of borrower – such as all exposures to corporate borrowers – are subject to the same capital requirement, regardless of potential differences in the creditworthiness and risk that each individual borrower might pose.

While the 1988 Accord was applied initially only to internationally active banks in the G10 countries, it quickly became acknowledged as a benchmark measure of a bank's solvency and is believed to have been adopted in some form by more than 100 countries. The Committee supplemented the 1988 Accord's original focus on credit risk with requirements for exposures to market risk in 1996. $\rangle \rangle$

The G10 group is made up of the following {\em eleven} countries: Belgium, Canada, France, Germany, Italy, Japan, the Netherlands, Sweden, Switzerland, the United Kingdom and the United States.

The Basel Committee was established in 1974 after the default of the German Bank 
{\em Bankhaus Herstatt}, with the purpose of making banking regulations more effective.

The cause of Herstatt's failure was its speculation in the foreign exchange markets.
After the collapse of the Bretton Woods system, in 1973, the management of the bank seriously underestimated the risk of free floating currencies. 

According to ref. \cite{failures} ``in September 1973, Herstatt became over-indebted as the bank suffered losses four times higher than the size of its own capital. The losses resulted from an unanticipated appreciation of the dollar. For some time, Herstatt had speculated on a depreciation of the dollar. Only late in 1973 did the foreign exchange department change its strategy. The strategy of the bank to speculate on the appreciation of the dollar worked until mid-January 1974, but then the direction of the dollar movement changed again. The mistrust of other banks aggravated Herstatt's problems.

In March 1974, a special audit authorised by the Federal Banking Supervisory Office
(BAKred) discovered that Herstatt's open exchange positions amounted to DM 2 billion,
eighty times the bank's limit of DM 25 million. The foreign exchange risk was thus three times as large as the amount of its capital \cite{Blei1984}. The special audit prompted the management of the bank to close its open foreign exchange positions.
When the severity of the situation became obvious, the failure of the bank could not be
avoided. 

In June 1974, Herstatt's losses on its foreign exchange operations amounted to
DM 470 million. On 26 June 1974, BAKred withdrew Herstatt's licence to conduct banking
activities. It became obvious that the bank's assets, amounting to DM 1 billion, were more than offset by its DM 2.2 billion liabilities.''

In international Finance, {\em Herstatt risk} refers to risk arising from the time delivery lag between two currencies. However, this risk was not the cause of the Herstatt
crisis.

\subsection{Scope of application}

Without entering into too many details, according to ref. \cite{basel2}, page 7, the Framework will be applied on a consolidated basis to internationally active banks, as the best means to preserve the integrity of capital in banks with subsidiaries by eliminating
double gearing. It is necessary to remark that the scope of application can still be
object of future negotiation. In 2003, the European Commission was eager to extend the proposal to the whole banking system, whereas the United States would have preferred to limit its scope only to internationally active banks. Moreover, some countries, such as India and China, declared that they do not wish to join the agreement \cite{Sole}

\subsection{The three pillars}

The Basel II framework rests on three pillars:

\begin{itemize}

\item
minimum capital requirements;

\item
supervisory review;

\item
market discipline.

\end{itemize}

\subsubsection{Minimum capital requirements}

Essentially, the agreement defines rules under which banks will be
able to compute the minimum capital requirements needed to perform risky
operations. According to the first pillar, three categories of risk are defined:

\begin{itemize}

\item
market risk;

\item
credit risk;

\item
operational risk.

\end{itemize}

Market risk and credit risk were already part of the 1988 agreement and the evaluation of market risk was already updated in 1996, with the possibility for banks of developing internal models. Therefore, the main innovations in \cite{basel2} concern the evaluation of credit risk and of operational risk. Operational risk is defined as the the risk of loss resulting from inadequate or failed internal processes, people and systems or from external events. This definition includes legal risk, but excludes strategic and reputational risk.

$\langle \langle$ First, Basel II improves the capital framework's sensitivity to the risk of credit losses generally by requiring higher levels of capital for those borrowers thought to present higher levels of credit risk, and vice versa. Three options are available to allow banks and supervisors to choose an approach that seems most appropriate for the sophistication of a bank's activities and internal controls. 

\begin{itemize}

\item
Under the {\em standardised approach} to credit risk, banks that engage in less complex forms of lending and credit underwriting and that have simpler control structures may use external measures of credit risk to assess the credit quality of their borrowers for regulatory capital purposes.

\item
Banks that engage in more sophisticated risk-taking and that have developed advanced risk measurement systems may, with the approval of their supervisors, select from one of two {\em internal ratings-based} (IRB) approaches to credit risk. Under an IRB approach, banks rely partly on their own measures of a borrowers' credit risk to determine their capital requirements, subject to strict data, validation, and operational requirements.

\end{itemize}

Second, the new Framework establishes an explicit capital charge for a bank's exposures to the risk of losses caused by failures in systems, processes, or staff or that are caused by external events, such as natural disasters. Similar to the range of options provided for assessing exposures to credit risk, banks will choose one of three approaches for measuring their exposures to operational risk that they and their supervisors agree reflects the quality and sophistication of their internal controls over this particular risk area.

By aligning capital charges more closely to a bank's own measures of its exposures to credit and operational risk, the Basel II Framework encourages banks to refine those measures. It also provides explicit incentives in the form of lower capital requirements for banks to adopt more comprehensive and accurate measures of risk as well as more effective processes for controlling their exposures to risk. $\rangle \rangle$

\subsubsection{Supervisory review}

As mentioned above, the degree of sofistication necessary for the evaluation of risks
is agreed upon by banks together with supervisory reviewers in order to ensure that bank management exercises sound judgement and that has set aside adequate capital to face these risks. This is the second pillar of the agreement.

$\langle \langle$ Supervisors will evaluate the activities and risk profiles of individual banks to determine whether those organisations should hold higher levels of capital than the minimum requirements in Pillar 1 would specify and to see whether there is any need for remedial actions. 

The Committee expects that, when supervisors engage banks in a dialogue about their internal processes for measuring and managing their risks, they will help to create implicit incentives for organisations to develop sound control structures and to improve those processes. $\rangle \rangle$

\subsubsection{Market discipline}

$\langle \langle$ Pillar 3 leverages the ability of market discipline to motivate prudent management by enhancing the degree of transparency in banks' public reporting. It sets out the public disclosures that banks must make that lend greater insight into the adequacy of their capitalisation. 

The Committee believes that, when marketplace participants have a sufficient understanding of a bank's activities and the controls it has in place to manage its exposures, they are better able to distinguish between banking organisations so that they can reward those that manage their risks prudently and penalise those that do not.
$\rangle \rangle$

\section{Basel II and physicists}

There are already reports from the Finance industry, that one of the major problems
in implementing Basel II requirements will be lack or shortage of skills and resources.
Research commissioned by Oracle found that the two biggest stumbling blocks for companies were a lack of access to relevant data (cited by 26 per cent of respondents) and a lack of skills (22 per cent) \cite{vnunet}.
Therefore, it is likely that the adoption of the Basel II agreement will foster both new jobs and new fields of research for physicists working on complex systems and  Econophysics. As mentioned above, after 1996, a new wave of physicists was hired by banks to work in risk management offices \cite{Milano2003} and a limited but significant number of new small enterprises led by physicists and active in the financial sector was born. Since then, a large amount of research efforts in Econophysics have been devoted to some aspects of market risk, including work on derivatives. However, in comparison to the pre-1996 situation, now:

\begin{itemize}

\item
there are several research groups of physicists all over the world working on problems inspired by Finance and Economics \cite{econophysics};

\item
some of these groups have joined larger collaboration at the national and international level. The European COST P10 Action {\em Physics of risk} \cite{COSTP10} is a good example of this trend;

\item
there is a number of journals devoted to the applications of Physics to Finance and Economics;

\item
many universities have started courses on Econophysics.

\end{itemize}

Therefore, new developments are envisaged in the analysis of both credit risk and
operational risk. Ref. \cite{basel2} devotes more than 100 pages to the evaluation of credit risk, moreover in the financial community there is an explosion of interest for credit risk and credit derivatives \cite{creditrisk}, whereas the author of this paper has found less than 20 preprints on credit risk in the Econophysics repository
{\tt www.unifr.ch/econophysics}. However, the approaches commonly used in Econophysics could be useful also for the analysis of credit risk. To illustrate this point, let us consider a recent working paper by Hui Chen, a PhD student in the Finance Program of the Graduate School of Business, University of Chicago \cite{chen}. As an instrument to price
bonds and loans using the standard present value approach, Chen suggests the use of 
{\em homogeneus continuous-time Markov chains} for the time evolution of credit-risk ratings. Standard \& Poor's, Moody's and other rating companies use a discrete set of 
credit risk assignments using labels such as AAA, Aaa, and similar. They also publish
one year transition probability matrices on which various commercial credit risk valuation programs are based. Most of these programs, such as CreditMetrics, CreditRisk+, CreditPortfolioView, use credit rating migration models based on discrete time Markov chains. In Chen's approach, a random variable in continuous-time, $x(t)$, describes
the present rating for a particular credit instrument and can assume any of a finite set of values (e.g. 8 values for S\&P or Moody's ratings). The time evolution of $x(t)$ is
based on the following conditional probability rule:
\begin{eqnarray}
P[x(s+t)& = & j|x(s)=i; (x(u): 0 \leq u < s)] = \nonumber \\
P[x(s+t)& = & j|x(s)=i]=P_{ij} (t).
\end{eqnarray}
In this way, one assumes that the embedded discrete-time process is a discrete-time
Markov chain. The time between state changes is a random variable with a memory-less distribution. The only con\-tinuous-time memory-less distribution is the exponential distribution. Therefore, the time between state changes in a continuos-time Markov chain is exponentially distributed. If we call $\lambda_i$ the inverse of the average waiting-time for transitions from state $i$ and we denote by $q_{ij}$ the one-step transition
probabilities of the embedded discrete Markov chain, we have the following relations:
\begin{eqnarray}
P_{ij}(\Delta t) & = & \lambda_i q_{ij} \Delta t, \, \, j \neq i \\
P_{ii}(\Delta t) & = & 1 - \lambda_i \Delta t. 
\end{eqnarray}
The meaning of the first equation is that the probability of the transition $i \to j$
in the time $\Delta t$ is the rate of such a transition, conditional on the initial state
$x(0)=i$ and multiplied by $\Delta t$. The second equation means that the probability of no transition taking place is 1 minus the probability that a transition takes place. Now,
the Chapman-Kolmogorov equation:
\begin{equation}
P_{ij} (t+ \Delta t) = \sum_{k} P_{ik}(\Delta t)P_{kj}(t),
\end{equation}
together with equations (2) and (3) yield in the limit $\Delta t \to 0$:
\begin{equation}
\frac{d P_{ij}(t)}{dt} = -\lambda_i P_{ij} (t) + \sum_{k \neq i} \lambda_i q_{ik} P_{kj}(t),
\end{equation}
a set of equations known as {\em Kolmogorov backward equations}. Chen goes on to give a simple recipe to derive coefficients in the Kolmogorov backward equations based on the one year transition matrices mentioned above.

The above example is only one possibility among many to apply stochastic methods familiar to physicists in the fields of credit and operational risk.

\section{Preliminary conclusions}

The preliminary conclusions of this paper are:

\begin{itemize}

\item
the Basel II agreement is likely to open new possibilites for physicists in the Finance industry;

\item
from the educational point of view, Econophysics courses should address issues related also to credit risk and operational risk and not only to market risk;

\item
as for research, a focus on applied risk management including work on credit and operational risks is likely to receive more attention from the financial community;

\item
in principle, physicists are well equipped to face the new challenges and opportunities determined by the Basel II agreement.

\end{itemize}

\section*{Final note}

The author welcomes
comments, suggestions and criticism. He can be reached at {\tt scalas@unipmn.it}.


\begin{thebibliography}{}

\bibitem{basel2}
Basel Committee on Banking Supervision, {\em International Convergence of Capital Measurement and Capital Standards: a Revised Framework}, 2004. The report can be downloaded from {\tt http://www.bis.org/publ/bcbs107.htm}.

\bibitem{release}
The full text of the press release is available from: \\ {\tt http://www.bis.org/press/p040626.htm}.

\bibitem{failures}
Basel Committee on Banking Supervision, {\em Bank Failures in Mature Economies},
Working Paper n. 13, April 2004.

\bibitem{Blei1984}
R. Blei, {\em Frueherkennung von Bankenkrisen, dargestellt am Beispiel der Herstatt Bank}, mimeo, Berlin, 1984.

\bibitem{Sole}
Il Sole 24 Ore, {\em Basilea 2. Guida pratica ai nuovi criteri per l'accesso al credito},
June 2004.

\bibitem{vnunet}
J. Mortleman, {\em Skills shortage creates Basel II compliance fears}, {\tt vnunet.com}
9 August 2004.

\bibitem{Milano2003}
In July 2003, in Italy, a meeting took place on physicists in Finance. A survey
showed that at least 100 physicists were working for financial institutions.
The proceedings of the meeting (mainly in Italian), are available from:
{\tt http://lagash.dft.unipa.it/talks.html}.

\bibitem{econophysics}
The main web sites devoted to this reserch field are: \\
{\tt http://www.econophysics.org}  \\
and {\tt http://www.unifr.ch/econophysics}.

\bibitem{COSTP10}
The web page of the COST P10 Action is: \\
{\tt http://gisc.uc3m.es/COST-P10/index.html}.

\bibitem{creditrisk}
The reader interested in credit derivatives could consult Vinod Kothari's portal: {\tt http://www.credit-deriv.com/}.

\bibitem{chen}
H. Chen, {\em A Continuous Time Compound Credit Rating Migration Model for Bond and Loan Valuations}, mimeo.

\end{thebibliography}
\end{document}